\documentclass{iaus}
\usepackage{graphicx}

\title[Magnetic fields in NGC\,4736] {Strong magnetic spiral pattern in a ringed galaxy NGC\,4736}

\author[K. T. Chy\.zy and R. J. Buta]  {Krzysztof T. Chy\.zy$^1$ \and Ronald J. Buta$^2$}

\affiliation{$^1$Astronomical Observatory of the Jagiellonian University, 
ul. Orla 171, 30-244 Krak\'ow, Poland \\[\affilskip]
$^2$Department of Physics and Astronomy, University of Alabama, USA
}

\pubyear{2007}
\volume{245}  \pagerange{119--126}
\date{?? and in revised form ??}
\jname{Formation and Evolution of Galaxy Bulges}
\editors{M. Bureau, E. Athanassoula \& B. Barbuy, eds.}
\begin{document}

\maketitle

\begin{abstract}
We present high sensitivity radio polarimetric (VLA) observations of a
galaxy with strong orbital resonances -- NGC\,4736. The total radio intensity
at 8.4\,GHz covers smoothly the whole galaxy bulge and reveals
a distinct ring of radio emission closely related to the ring morphology 
visible in infrared, CO and H$\alpha$ emission. However, the magnetic 
field reveals a very coherent spiral pattern. The magnetic
field vectors are crossing the inner starbursting ring, the dust lanes within
the ring and other rather circularly shaped features visible
in other gas traces. Either the magnetic field uncovers the pattern of gas
motions not seen in other spectral ranges, or the spiral
magnetic field is of a pure dynamo origin, ignoring the ringed
morphology of the galaxy. 

\end{abstract}

\firstsection               \section{Radio polarimetric observations of a ringed galaxy}

How the galactic magnetic field can operate in early Hubble-type galaxies 
and in the presence of strong orbital resonances is as yet unexplored. 
Using the VLA we investigated this problem and observed 
NGC\,4736 (M\,94) -- the nearest and largest
galaxy showing a ringed morphology (Buta \& Combes \cite{buta96}). 
Its inner and outer rings are believed to
coincide with Lindblad resonances of a triaxial bulge, the inner disk
(oval distortion) and/or the nuclear minibar. The giant inner ring 
located at a distance of about 47" (1.1 kpc) from the (LINER) nucleus,
is a region of intense star formation which dominates the UV, H$\alpha$,
infrared and CO images (Wong \& Blitz \cite{wong01}). 

\begin{figure}
  \includegraphics[width=5.4in]{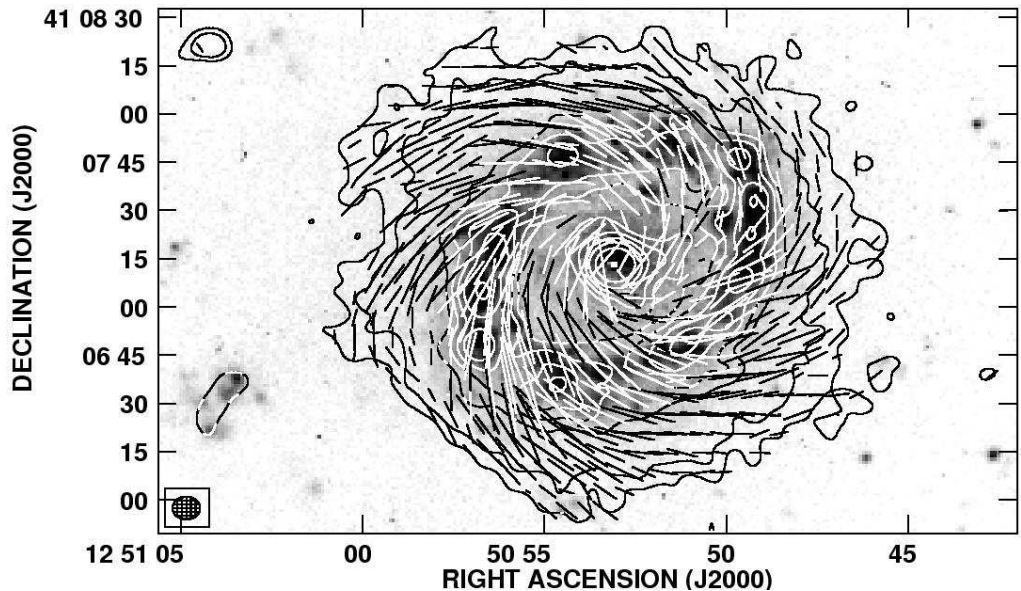}
  \caption{The total radio intensity contour map of NGC\,4736 at 8.4\,GHz
at 8"x7" resolution with observed magnetic field vectors of the polarized 
emission overlaid on the H$\alpha$ image (from Knapen et al. \cite{knapen03}). 
Contours are at (3, 8, 32, 50, 80, 128, 256)\,$\times6\,\mu$G/b.a. 
Vectors of length of 10" correspond to polarized intensity of $26\,\mu$G/b.a.
}
\label{f:tp}
\end{figure}

We observed NGC\,4736 with the VLA in its D array at 8.4 and 4.6\,GHz.  
The observed total radio intensity covers smoothly
the whole galaxy bulge and reveals a distinct
ring of radio emission (Fig.~\ref{f:tp}). The radio ring is closely
related to the ring morphology visible in other spectral ranges,
but the radio polarized emission at 8.4\,GHz  
seems to be not clearly associated with H$\alpha$ emitting gas,
or other ISM phases. Instead, it reveals a very coherent spiral 
pattern of the regular magnetic field (Fig.~\ref{f:tp}). 

Surprisingly, the prominent inner ring does not tune the magnetic field
vectors. Two magnetic spiral arms of constant pitch angle of about $45^o$
emerge from the centre of NGC\,4736 and cross the
inner ring. This is opposite to what is observed in grand-design spiral galaxies
where the regular magnetic field typically follows the shape of nearby 
gaseous spiral arms, even if a galaxy is perturbed (Chy\.zy et al. 
\cite{chyzy07}).

Between the arms, and inside the ring, the regular field of NGC\,4736 
has a smaller pitch angle. This seems to be related to a central minibar
extending along the galaxy's minor axis.
The magnetic field vectors also cross the NW long and strong dust lane, 
visible in optical images. A spiral network of dust
armlets in the central part of the galaxy (probably of acoustic origin) 
seems also to be not associated with the
magnetic spiral -- the magnetic vectors have typically
larger pitch angles, by at least $20^o$. Nevertheless, the magnetic field
may still help in channelling the surrounding gas and dust into an active
LINER nucleus.

\section{Magnetic spiral}

The discovered spiral structure of the magnetic field in NGC\,4736
seems to contradict the galaxy's ringed morphology. If the magnetic
field is totally frozen into the interstellar plasma then the observed
magnetic vectors may reveal the pattern of the galactic gas motions and
shocks which is not yet discerned in other spectral ranges. On the other
hand, and more likely, the spiral magnetic field can be
of pure dynamo origin. In this case the magnetic field and the interstellar
turbulence must be strong enough to resist the local gas flows and
cross the star-forming ring imperceptibly. In each case NGC\,4736 seems
to be a key object to study the resonant dynamics and a dynamo process 
of the  magnetic field generation without spiral geometry of density waves.

\begin{acknowledgments}
This work was supported by the Polish Ministry of Science and Higher
Education, grant 2693/H03/2006/31.
\end{acknowledgments}

\end{document}